# Determination of the ferrimagnetic structure of monoclinic $TbFe_2D_{4.2}$ deuteride and its evolution versus temperature and high magnetic field

V. Paul-Boncour[1], O. Isnard[2]

[1] *Université Paris-Est Créteil, CNRS, ICMPE, UMR7182, F-94320 Thiais, France*

[2] *Université Grenoble Alpes, Institut Néel, CNRS, BP166X, 38042 Grenoble Cédex 9, France*

* Corresponding author: valerie.paul-boncour@cnrs.fr

*This paper is dedicated to the memory of Maurice Guillot, who contributed to this study by high magnetic field measurements.*

**Abstract**

The magnetic properties of monoclinic $TbFe_2D_{4.2}$ have been investigated by combining different magnetic measurements, including those involving a high pulsed magnetic field, with neutron diffraction experiments versus temperature and applied field. $TbFe_2D_{4.2}$ displays a ferrimagnetic ground state with $M_{Tb}$ = 8.2(2) $\mu_B$ and $M_{Fe}$ = 2.1(2) $\mu_B$. At 4.2 K a large magnetic anisotropy related to the Tb moments is observed, as saturation is not reached at 50 T. In addition, two metamagnetic transitions are observed for fields of 3 and 29 T. Upon heating, the Tb moment decreases progressively whereas the Fe moment remains constant up to 150 K. At 160 K both Tb and Fe moments sharply decrease, and a 0.38% contraction of the cell volume is observed due to an itinerant electron metamagnetic behavior of the Fe sublattice. This transition temperature is 76 K higher than that of $YFe_2D_{4.2}$ due to an increase in cell volume, as well as exchange interactions between the Fe and Tb moments. The deuteride becomes paramagnetic above 220 K.

Keywords : Laves Phase, deuteride, Neutron diffraction, Ferrimagnetism

## 1. Introduction

$RFe_2$ Laves phase compounds have been attracting large interest for their magnetic properties, particularly for their giant magnetostrictive and magnetoelastic properties near room temperature [1, 2]. Such strong magnetostrictive effects are explained by the combination of the large magnetic anisotropy of the rare earth elements and the high Curie temperature related to Fe. The largest magnetostrictive strain was observed for $TbFe_2$ (2630 ppm), which properties can be tuned by many chemical substitutions on both Tb and Fe sites by other rare earth and transition elements respectively [3-9]. $Tb_{0.3}Dy_{0.7}Fe_y$ ($y$ = 1.9-1.95) compound, called Terfenol-D (Ter for Terbium, Fe for iron, Nol for Naval Ordonance Lab and -D for Dysprosium), has been developed by Clark et al [10] in USA for naval sonar systems. Its synthesis and magnetic properties have been studied by various methods [11-17]. It is now commercialized and used in many applications such as magnetomechanical sensors, actuators, acoustic and ultrasonic transducers due to its high energy density and wide bandwidth capabilities [18-24].

Beside chemical substitution, hydrogen insertion was also tested to tune the magnetic properties of Terfenol-D through the modification of the cell volume and electronic properties [25]. The structural and magnetic properties of $TbFe_2H(D)_x$ hydrides and deuterides were also investigated showing an interesting evolution versus the H(D) content [26-34]. A detailed structural analysis is reported in [35]. Most $TbFe_2$ hydrides and deuterides are cubic up to $x$ = 3, except a tetragonal superstructure for $x = 2$, and they present monoclinic distortions for $3.6 \leq x \leq 4.2$. Cubic deuteride was obtained again for larger D content ($x = 4.5$). According to previous works, the $TbFe_2H(D)_x$ compounds are ferrimagnetic [26, 27]. The molecular field, obtained by the modeling of the magnetization curves with a Brillouin function, decreases with H content showing a reduction of the Tb-Fe interactions [26, 27]. A reduction of the hyperfine field transferred from the rare earth toward the Fe atoms was also observed versus H content in [27]. The saturation magnetization of $TbFe_2H_x$ hydrides at low temperature decreases versus H content. A rhombohedral distortion associated to a discontinuous and sharp reduction of the magnetization is observed for $x \geq 2.65$ [27]. In addition, the increase of the slope of the linear part of magnetization was attributed to the anisotropic distortion. Kulshreshtha et al. [28] observed a spin reorientation of the Fe moments upon H absorption using $^{57}$Fe Mössbauer spectroscopy, from (111) direction for $TbFe_2$ to probably (100) for the hydrides. They have also observed that $TbFe_2H_{4.8}$ is cubic and paramagnetic at 78 K. Zaikov et al [36] investigated the magnetic properties of several hydrogen-amorphized $R$Fe$_2$ hydrides including $R$=Tb. They observed an increase in the Fe-Fe and a decrease in the $R$-Fe exchange interaction energies.

In a previous study [37], we have found that according to laboratory scale X-ray diffraction data $TbFe_2H_{4.2}$ and $TbFe_2D_{4.2}$ are both monoclinic described in a $C2/m$ space group with a larger cell volume for the hydride compared to the deuteride. The structure of the deuteride $TbFe_2D_{4.2}$ was then fully solved by neutron diffraction experiments showing a lowering of the crystal symmetry to monoclinic $Pc$ space group with a doubling of the $b$ cell parameter [35] as in $YFe_2D_{4.2}$. $YFe_2H_{4.2}$ and $YFe_2D_{4.2}$ both display a first-order ferromagnetic-antiferromagnetic (FM-AFM) transition of the Fe sublattice [38, 39]. A giant (H/D) isotope effect was observed with an increase of $T_{FM\text{-}AFM}$ from 84 K for the deuteride to 131 K for the hydride which was attributed to a magnetovolume effect [40]. The FM-AFM transition is accompanied by a significant negative cell expansion and a negative change of the magnetic entropy upon heating [38] characteristic of an itinerant metamagnetic behavior of the Fe sublattice. This $T_{FM-AFM}$ transition temperature can be tuned by the substitution of Y by another rare earth element as observed in several $Y_{1-y}R_yFe_2H(D)_{4.2}$ compounds ($R$= Pr, Gd, Tb, Er) [41-44] as well as applied pressure [39, 45, 46].

The study of $Y_{1-y}Er_yFe_2(D)H_{4.2}$ has shown that $T_{FM-AFM}$ decreases linearly versus Er content and is systematically lower for the deuteride compared to the hydride for a given Er content. This was attributed to the cell volume contraction upon Er substitution and D for H replacement. In addition, two types of metamagnetic transitions were observed in $Y_{1-y}Er_yFe_2(D)H_{4.2}$ compounds [41-43]. The first one at low temperature and around 8 T is due to a forced spin reorientation from a ferrimagnetic toward a ferromagnetic state of the Er sublattice and is not accompanied by any cell volume anomaly. The second one above $T_{FM-AFM}$ is related to the spin reorientation of the Fe moments from an AFM to a magnetic FM state accompanied by a cell volume contraction. Another study on $Y_{0.9}Tb_{0.1}Fe_2D_{4.2}$ has revealed a multistep spin reorientation of the Tb moment at low temperature and an increase of $T_{FM-AFM}$ compared to $YFe_2D_{4.2}$ [40, 47] Concerning $TbFe_2D_{4.2}$, a cell volume contraction was observed around 160 K, but this transition is not visible in the magnetization curves, which display a continuous decrease of the magnetization related to a ferrimagnetic behavior. Few differences were also observed on the magnetization curves versus temperature of $TbFe_2D_{4.2}$ and $TbFe_2H_{4.2}$ at low temperature and above 150 K. But in this previous work the experimental results were not fully analyzed, and some questions remained unsolved such as the magnetic behavior of each Tb and Fe sublattices and how they are correlated.

For this purpose, we have investigated in detail the magnetic properties of $TbFe_2D_{4.2}$ by combining magnetic measurements at low and high magnetic field, including pulsed magnetic field up to 50 T, with different neutron diffraction experiments versus temperature and applied field. The experimental results will be discussed in comparison with other studies concerning $Y_{1-x}R_xFe_2(H,D)_{4.2}$ compounds.

## 2. Experimental methods

The $TbFe_2$ alloy was synthesized by induction melting of the pure elements followed by a 3-week annealing treatment at 1100 K under secondary vacuum ($2.10^{-5}$ mbar). A small excess of Tb (5%) was added to compensate for any loss due to Tb oxidation during the heat treatment.

The alloy composition was measured by electron probe microanalysis (EPMA) on a mirror-polished sample. Details of the synthesis and analysis of the samples used below are given in ref [35].

The $TbFe_2D_x$ deuteride was prepared by a Sieverts method using deuterium gas using 7.5 g of alloy. The final pressure was 2 bars, and the calculated D content was 4.2(1) D/f.u.. The sample was quenched in liquid nitrogen and slowly heated to room temperature under air to passivate the surface and avoid D desorption at room temperature, which can occur after several days.

The neutron powder diffraction (NPD) experiments were performed at three different neutron centers as described below. The NPD patterns have been recorded at 2 and 310 K on the high resolution 3T2 diffractometer at the Laboratoire Léon Brillouin (LLB, CEA, Saclay, France) with $\lambda$= 1.229 Å (15 ° < 2θ < 125 ° and a step of 0.05 °). The NPD patterns were also measured between 2 and 300 K in an ILL orange cryostat on the high intensity two axis diffractometer D1B ($\lambda$ = 2.523 Å, 10.5 °< $2\theta$< 90.3°, step= 0.2° ) at the Institut Laue Langevin (ILL, Grenoble, France). Furthermore, NPD patterns were measured at different temperatures under an applied field up to 12 T using a vertical field cryo-magnet on E6 focusing diffractometer at Helmoltz Zentrum Berlin (HZB, Berlin, Germany). The selected wavelength was 2.447 Å (3 ° < 2θ < 113 ° with a step of 0.15 °). The process to prepare the sample for these measurements is detailed for $Y_{0.9}Tb_{0.1}Fe_2D_{4.3}$ in ref. [44]. All the NPD patterns were refined using Rietveld refinement with the Fullprof code [48].

Magnetic measurements were performed using a Physical Properties Measurement System (PPMS) from Quantum Design. Additional measurements were performed at selected temperatures between 4.2 K and 300 K under continuous field ($\mu_0 H_{max}$ = 23 T) by an extraction method at the Laboratoire National des Champs Magnétiques Intenses in Grenoble (LNCMI-G). A magnetic curve at 4.2 K under pulsed magnetic fields up to 60 T was recorded at the Dresden High Magnetic Field Laboratory (DHMFL) in Germany. This curve was scaled using the LNCMI-G data at 4.2 K up to 23 T.

## 3. Results and discussion

The crystal structure of $TbFe_2D_{4.2}$ at 300 K was solved by neutron diffraction and is described in [35]. The compound crystallizes in a monoclinic structure described in *P*c space group with $a$ = 5.5171(2) Å, $b$ =11.5061(5) Å, $c$= 9.4382(3) Å, $V$ = 506.19(3) Å$^3$ and $\beta$= 122.342(2)°. Due to the sensitivity of neutron scattering to deuterium, this crystal structure is considered to be more reliable than the average monoclinic *C*2/*m* structure found from XRD only [64]. The D atoms occupy only 18 tetrahedral interstitial sites (13 [$Tb_2Fe_2$] and 5 [$TbFe_3$] sites) among 64 possible sites. The ordering of D atoms in few interstitial sites is related to the geometric constraints such as the Westlake [49] and Switendick [50] criteria but also the number of bonds between D and Fe or Tb atoms.

*3.1.* Magnetic measurements

The magnetic curves $M(T)$ of $TbFe_2D_{4.2}$ measured with the PPMS under applied field of 0.03 and 1.5 T are presented in Fig. 1*a*. The magnetization curve at 0.03T progressively decreases to a minimum at 170 K and slightly increases above this temperature. Furthermore, a small bump is observed between 170 and 240 K with a maximum at 220 K for an applied field of 1.5 T. The origin of this magnetic contribution will be discussed later in relation to the neutron diffraction results.

The isothermal magnetization curves measured versus applied field $M(\mu_0 H)$ are plotted in Fig. 1b (PPMS), Fig. 1c (LNCMI-G) and with a pulsed magnetic field at 4.2 K in Fig. 1d (DHMFL). A remarkable feature of the isothermal curves is the absence of saturation even to high magnetic fields of 23 and 50 T. On the contrary, a large slope is observed in $M(\mu_0 H)$ curves, a slope of 0.11(1) $\mu_B$/f.u./T being deduced at 2 K. Such behavior may reflect the ferrimagnetic nature of the $TbFe_2D_{4.2}$ compound at low temperature and the progressive field induced rotation of the magnetic moments towards the applied magnetic field. This slope is larger than that of the parent $TbFe_2$ compound which displays a slope of 0.066(5) $\mu_B$/f.u./T at 5 K. Concerning $TbFe_2D_{4.2}$ several metamagnetic transitions are observed. From the derivative of the magnetization curves of $TbFe_2D_{4.2}$ at 4.2 K and 5 K (figs. 1b, 1c and 1d) two transition fields $\mu_0 H_{Trans}$ are observed at 3 T and 29 T respectively. In Fig. 1b the metamagnetic behavior at low magnetic field is observed in the $M(\mu_0 H)$ curves between 5 and 60 K. The transition field $\mu_0 H_{Trans}$, obtained by the derivative of the curves, decreases from 3 to 1.1 T upon heating. Above 100 K this first transition field is no longer observed.

The spontaneous magnetization deduced from the $M(\mu_0 H)$ curve of $TbFe_2$ at 5 K between 2.5 and 5 T reaches 5.2(1) $\mu_B$/f.u.. Assuming a collinear ferrimagnetic structure and a Tb moment of 9.5 $\mu_B$/atom (free ion value), the deduced Fe moment is 2.15 $\mu_B$/atom. For $TbFe_2D_{4.2}$ at 5 K measured on the PPMS reveals two different regions with spontaneous magnetization at 2.5(1) and 3.3(1) $\mu_B$/f.u. respectively (Fig. S1, supplementary materials). This means either a reduction of the Tb moment by crystal field effects or a distribution of the orientation of the Tb moments. If the value of the Fe moments is the same as in the parent compounds, the Tb moment deduced from the spontaneous magnetization at 5 K for a collinear ferrimagnetic arrangement would be respectively of 6 and 7.6 $\mu_B$/f.u. depending on the selected range (Fig. S1). The refinement of the NPD patterns at low temperatures will be useful to interpret these results.

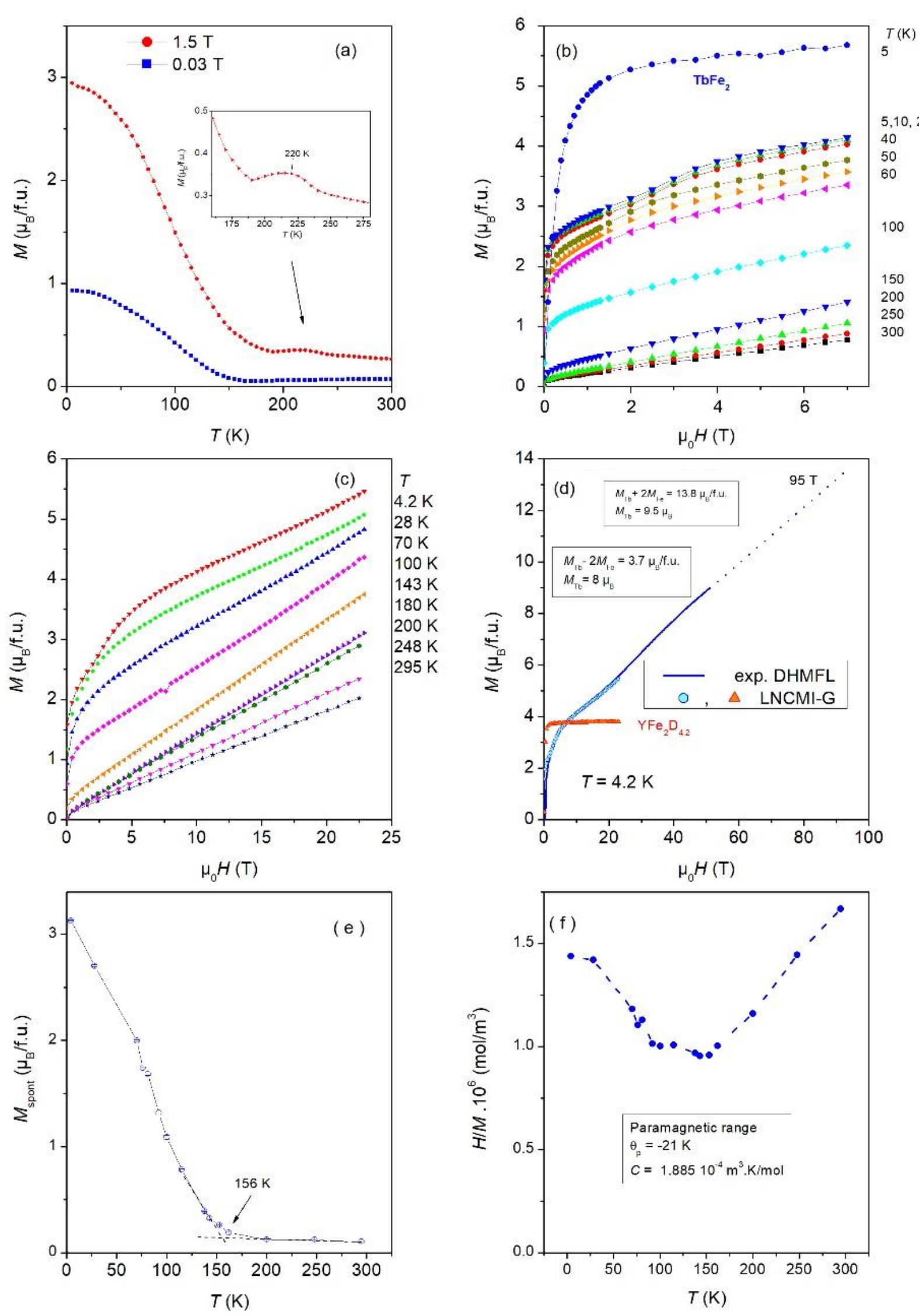

**Figure 1**: Magnetization curves of $TbFe_2D_{4.2}$ (a) $M(T)$ measured at 0.03 and 1.5 T with a PPMS device (zoom of 1.5 T curve in inset), (b) $M(\mu_0 H)$ curves measured at selected temperatures with PPMS device ($TbFe_2$ added for comparison), (c) $M(\mu_0 H)$ curves measured at LNCMI-G (d) $M(\mu_0 H)$ measured at 4.2 K using a pulsed magnetic field at DHMFL (curves of $TbFe_2D_{4.2}$ and $YFe_2D_{4.2}$ measured at LNCMI-G added for comparison), (e) spontaneous magnetization and (f) inverse of the magnetic susceptibility deduced from the slopes of the $M(\mu_0 H)$ curves of fig.1c, fit of the linear part with a Curie Weiss law.

Assuming a moment of 9.5 $\mu_B$/f.u. for Tb and a total magnetization of 4.3 $\mu_B$/f.u. for the Fe sublattice (as observed for the $TbFe_2$ alloy) one can expect to reach a saturation value 5.2 $\mu_B$/f.u. in a collinear ferrimagnetic state and of 13.8 $\mu_B$/f.u. in a ferromagnetic structure at large magnetic field. A linear interpolation of the DHMFL $M(\mu_0 H)$ curve at 4.2 K yields a field of at least 95 T to reach this value (Figure 1d).

The linear part of the $M(\mu_0 H)$ curves of fig. 1c were fitted between 10 and 20 T to obtain the thermal variation of the spontaneous magnetization $M_{spont}$ by the intercept at zero field and the the slope d$M$/d$H$ (Fig. 1e). The thermal variation of $M_{spont}$ indicates a transition temperature near 156 K, but the magnetization remains constant above 200 K. The analysis of the magnetic structure by NPD is necessary to interpret these data. The inverse of the magnetic susceptibility displays a hyperbolic form characteristic of a ferrimagnetic compound. The linear fit of the inverse of the magnetic susceptibility above 150 K yields a negative paramagnetic temperature $\theta_p$ = -21(1) K revealing the existence of dominant antiparallel moments as expected for ferrimagnetic coupling. The calculated Curie constant is 15(1) $cm^3$.K /mole. This value corresponds to the sum of the Tb and Fe Curie constants. Assuming an effective moment of 9.72 $\mu_B$ for Tb yields to C = 11.8 emu/mole. The difference of 3.2 emu/mole can be attributed to the contribution of the Fe moments.

3.2 Neutron diffraction experiments

a) *Magnetic structure at 1.5-2 K*

The NPD patterns measured at 1.5 and 310 K on D1B diffractometer (ILL) are compared in Figure 2a. The difference pattern between these two temperatures (Figure 2b), refined with a pattern matching method, highlights the existence of several magnetic Bragg peaks which can be indexed in the same monoclinic cell than the nuclear one ($a$ = 5.500(1) Å, $b$ = 11.515(2) Å, $c$ = 9.428(1) Å, $\beta$ = 122.004(1)°), indicating a $k$ = (000) propagation vector.

The NPD pattern of $TbFe_2D_{4.2}$ measured at 1.5 K with D1B diffractometer was refined using the Rietveld method in a monoclinic cell ($Pc$ space group) with $a$ = 5.516(1) Å, $b$ = 11.534 (2) Å, $c$= 9.394(2) Å, $\beta$ =122.30(1). A joint Rietveld refinement was performed with the full patterns measured at 1.5 K on D1B diffractometer and at 2 K on 3T2 diffractometer (Figure 2c and 2d respectively). The reliability factors are indicated in the legend. The atomic positions and occupation number of Tb, Fe and D atoms were fixed according to the nuclear structure in the paramagnetic state refined at 300 K [35]. The magnetic contribution to the NPD pattern is well refined considering a collinear ferrimagnetic structure with the Tb and Fe moments parallel to the (*a*, *c*) plane.

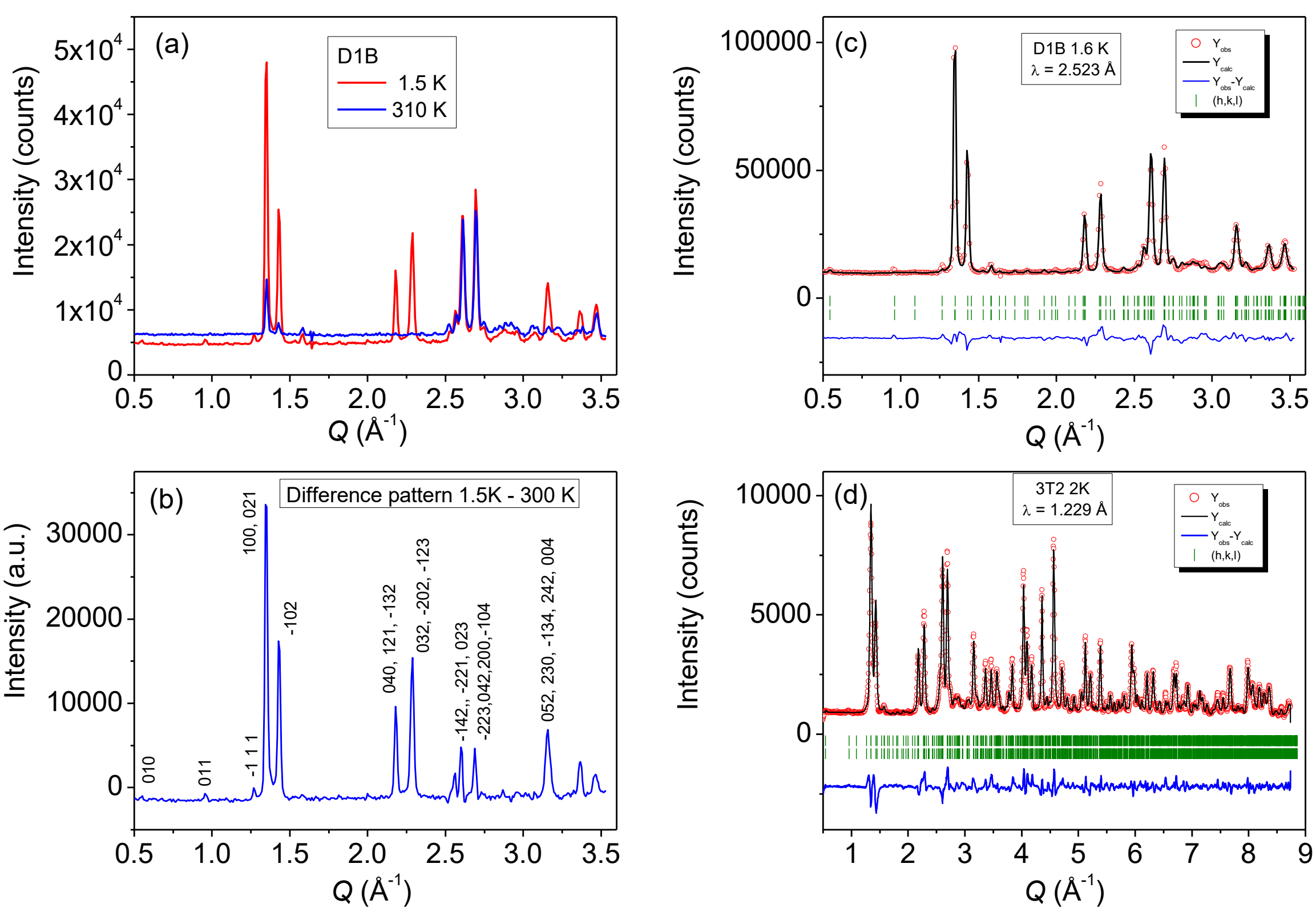

**Figure 2**: Neutron diffraction patterns of $TbFe_2D_{4.2}$ (a) at 1.5 and 310 K measured on D1B diffractometer (ILL) ($\lambda$ = 2.523 Å), (b) corresponding difference pattern with (hkl) indexations (c) NPD pattern measured on D1B (ILL) at 1.5 K and (d) NPD pattern measured on 3T2 (LLB) at 2 K ($\lambda$= 1.229 Å) refined together by a Rietveld method. The reliability factors are respectively: $R_p$: 16.0, $R_{wp}$:14.7, $R_{Bragg}$: 12.8, $R_{Mag}$: 6.78 for the D1B pattern and $R_p$: 12.1, $R_{wp}$: 3.9, $R_{Bragg}$: 6.32 and $R_{Mag}$: 8.80 for the 3T2 pattern.

A refinement with a component along the ***b*** axis was tested, but it doesn't improve the quality of the fit. The moment magnitude was fixed with the same values for all Tb and Fe moments respectively and the refinement yields $M_{Tb}$ = 8.2(2) $\mu_B$/Tb et $M_{Fe}$ = 2.1(2) $\mu_B$/Fe. The angle between the *a* and *c* axis is 87(1) ° close to 90 ° indicating that the moments are almost perpendicular to the **c**-axis. The Tb moment is smaller than the free ion value (9.5 $\mu_B$), whereas the Fe moment is slightly smaller than for $TbFe_2$ ($M_{Fe}$ =2.15 $\mu_B$/atom). $M_{Fe}$ is larger than that measured for $YFe_2D_{4.2}$ ($M_{Fe}$ = 1.9 $\mu_B$) due to the exchange interactions between Fe and Tb moments.

*b) Evolution of the magnetic structure versus temperature at zero field*

The evolution of the NPD patterns measured versus temperature on D1B diffractometer (ILL) is presented in Fig. 3 as a 3D plot.

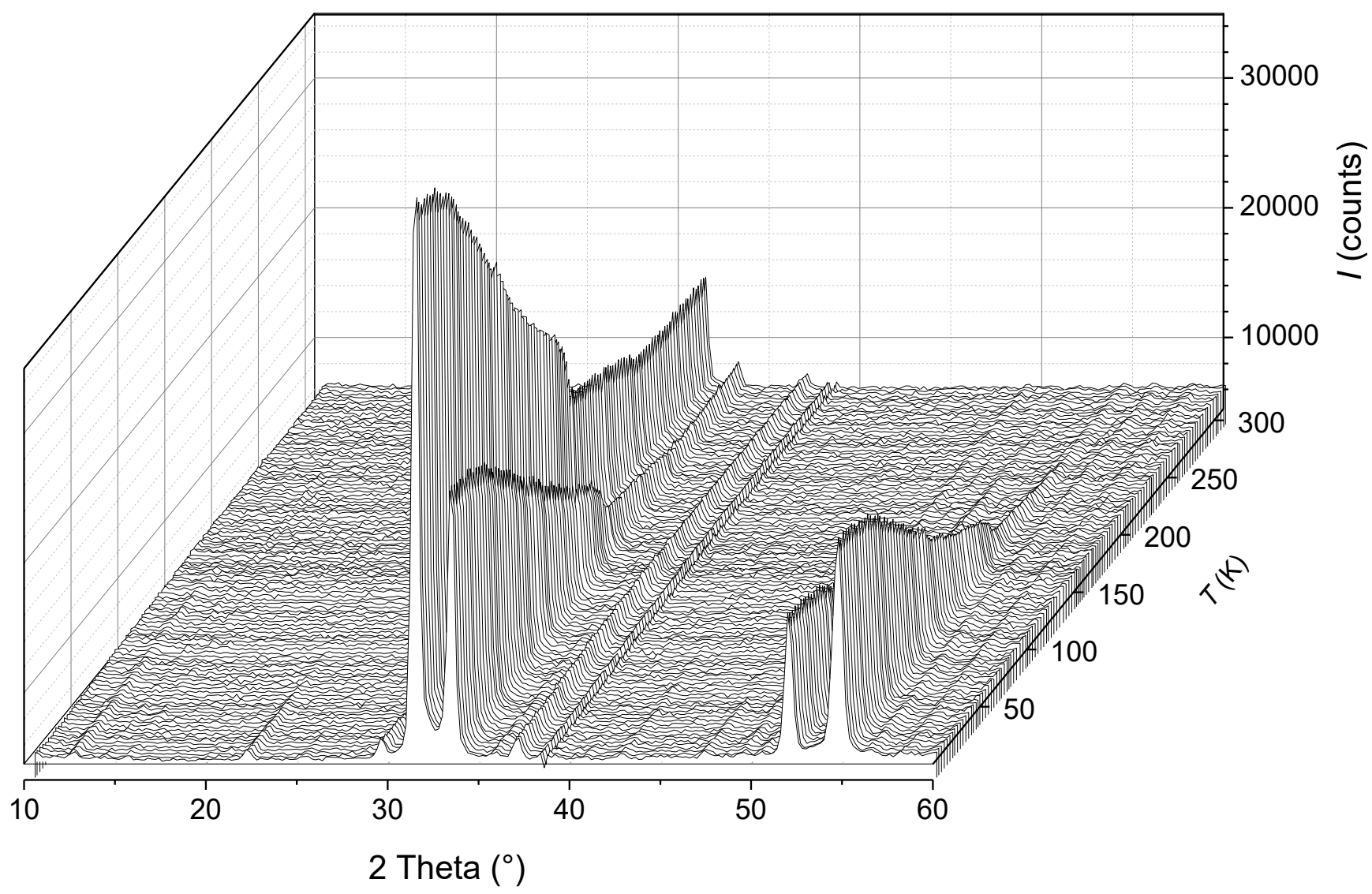

**Figure 3** : 3D plot showing the thermal evolution of the low angle part of the NPD patterns of $TbFe_2D_{4.2}$ measured on D1B (ILL) with λ = 2.52 Å.

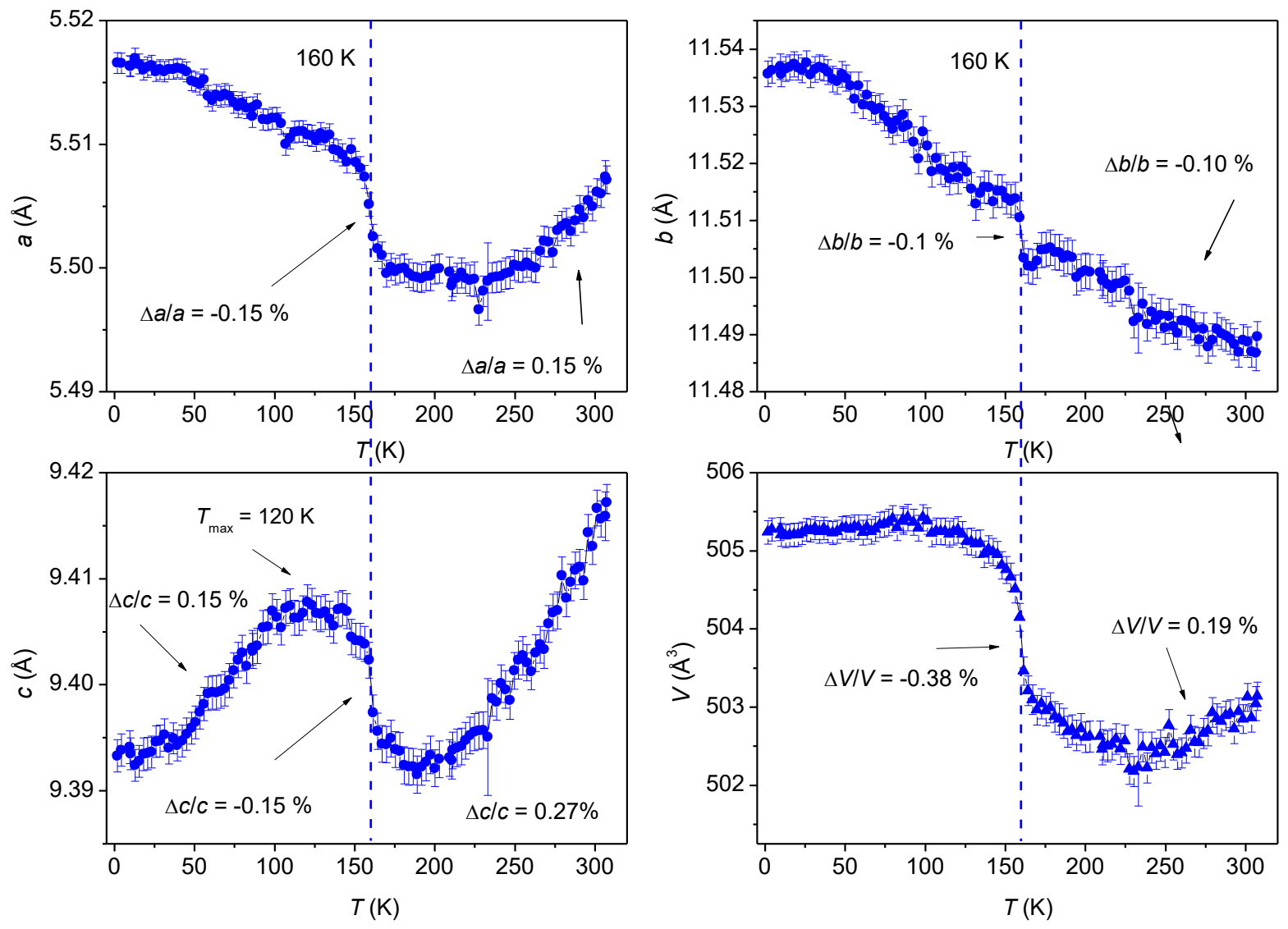

**Figure 4** : Evolution of the monoclinic lattice parameters as a function of temperature for the $TbFe_2D_{4.2}$ deuteride obtained from the Rietveld refinement of the NPD patterns (D1B, ILL) in *Pc* space group. The relative cell parameter variations between 150 and 170 K (magnetovolume contraction) and between 210 and 310 K (competition between thermal expansion and reduction of the monoclinic distortion) are reported in the plot.

Below 220 K, the decrease of the peak intensity can be related to the change of magnetic peak intensity. Above this temperature the intensity of the peak decreases slowly and can be attributed to the nuclear contribution as expected in a paramagnetic range.

All the NPD patterns measured on D1B were refined using the Rietveld method with the same nuclear and magnetic structure as at 2 K. The evolution of the lattice parameters and the cell volume as a function of temperature is presented in Figure 4. Several domains can be observed related to different magnetic behaviors. Between 2 and 120 K, the *c* parameter increases of 0.15 %, whereas the *a* and *b* parameters decrease upon heating. This variation can reflect the ferrimagnetic behavior between Fe and Tb along the ***c***-axis. A sharp transition with negative expansion is observed around 160 K with a cell volume contraction of 0.38%. The cell volume contraction is slightly smaller than observed in $YFe_2D_{4.2}$ deuteride (0.5 %). Note that this negative expansion is anisotropic, being of -0.15 % for *a* and *c* cell parameters and only -0.10 % for the *b* cell parameter. This can be due to the main orientation of the Fe magnetic moments in the (*a*, *c*) plane.

A minimum of the cell parameters is observed at 190 K for *c* and 230 ±5 K for *a* and *V*. The cell parameters *a*, *c* and *V* increase again upon heating, whereas *b* decreases continuously up to 300 K. Previous studies on $YFe_2D_{4.2}$ have associated this minimum to a transition of the Fe sublattice from an antiferromagnetic to a paramagnetic state [47]. The analysis of the refined magnetic parameters will be useful to determine if this transition has the same origin.

The anisotropic variation of the *a*, *b* and *c* parameters above 220 K, particularly the decrease of the *b* parameter, can be explained by the reduction of the monoclinic distortion which is in competition with the thermal expansion. In previous work [35] it was observed that $TbFe_2D_{4.2}$ presents an order-disorder structural transition from a monoclinic to a cubic structure around 340-400 K, a transition induced by the loss of the long range order of deuterium atoms. The monoclinic structure is related to a long-range order of the D atoms. Upon heating the D atoms start to move more easily from one site to the neighboring one, and the long-range order disappears above a critical temperature denoted $T_{O\text{-}D}$. The cubic structure is an average structure with D atoms located statistically into [$Tb_2Fe_2$] and [$TbFe_3$] interstitial sites. It was observed using partial distribution function (PDF) that the local order of D atoms in $YFe_2D_{4.2}$ is preserved inside the unit cell [51], a similar behavior is expected for $TbFe_2D_{4.2}$. The monoclinic distortion from the $MgCu_2$ type cubic structure yields a contraction of the *a* and *c* parameters and an expansion of the *b* parameter. Therefore, the reduction of the monoclinic distortion increases *a* and *c* and reduces *b* as observed in Figure 4.

To understand the evolution of the magnetic structure, one can observe at first the thermal variation of the main magnetic peaks before examining the results of the Rietveld refinement.

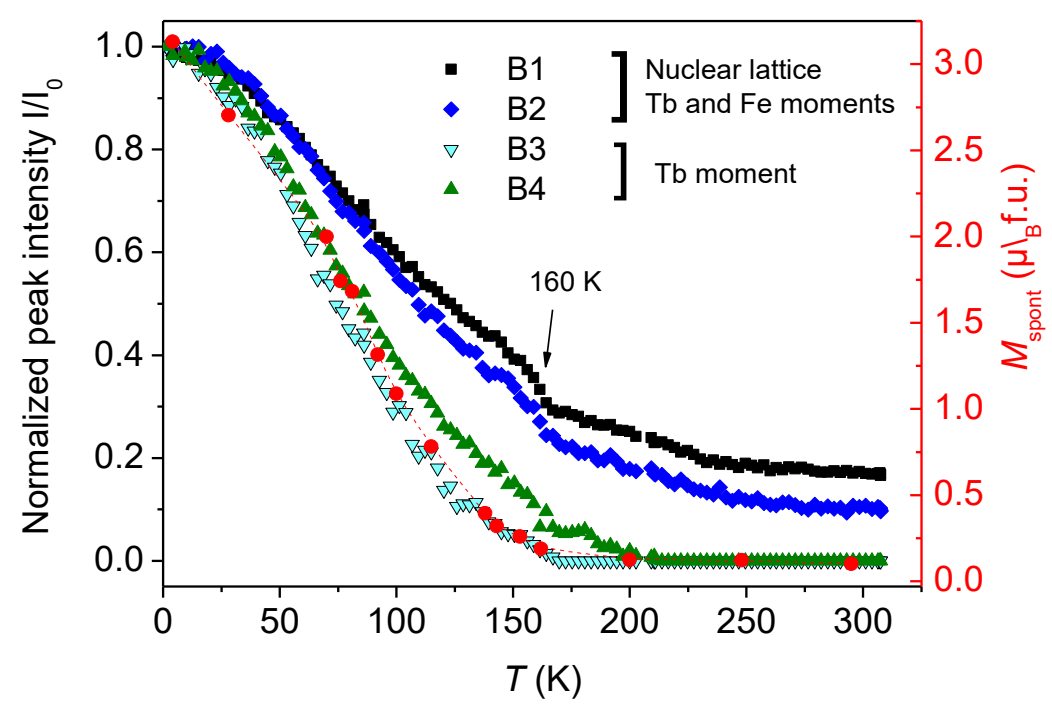

**Figure 5**: Evolution of the normalized intensity of four selected Bragg peaks as a function of temperature for the deuteride $TbFe_2D_{4.2}$. The spontaneous magnetization $M_{spont}$ (Fig. 2e) has been added for comparison, the corresponding scale is given in the right axis.

The normalized intensity of four Bragg peaks versus temperature is reported in Figure 5. These peaks are identified with their Bragg indices. The two first peaks near $Q$ = 1.35 Å$^{-1}$ ((021), (100)) and $2\theta$ = 1.41 Å$^{-1}$ (-102), denoted B1 and B2 respectively, contain both nuclear and magnetic intensities as they are still observed in the paramagnetic range. The Bragg peak at $Q$ = 2.18 Å$^{-1}$ (-132), (040), (121)) denoted B3 is purely magnetic whereas the B4 Bragg peak at $Q$ = 2.29 Å$^{-1}$ ((-123) and (-202)) contains mainly a magnetic contribution and a very weak nuclear one (see Figure 1a). The Rietveld analysis shows that these two peaks are mainly related to the Tb moments (observed when Fe moments are fixed to zero in the simulation, Suplementary materials, Figure S2). Therefore, the variation in their intensity reflects the variation of the Tb moments. One can notice in Figure 5 that the B3 peak intensity follows well the variation of the spontaneous magnetization, and it becomes equal to zero above 170 K. The B4 peak intensity is visible up to 210 K. A sharper decrease of intensity is observed in the B1 and B2 peaks at 160 K, precisely the temperature where all the cell parameters are contracted.

To obtain the thermal variation of the Tb and Fe moments, the NPD patterns measured on D1B were refined with the Rietveld method in a spherical description of the magnetic cell with the Tb and Fe moments perpendicular to the ***b*** axis as for the refinement at 2 K. The Fe and Tb moments were constrained to have the same magnitude for all the Tb and Fe sites respectively, to avoid a divergence of the fit. The variation of the Tb and Fe moments ($M_{Tb}$ and $M_{Fe}$) as well as the φ angle between the moment orientation and the ***c***-axis are plot in Figure 6. $M_{Tb}$

progressively decreases from 8.0 to 3.0 $\mu_B$/atom whereas $M_{Fe}$ remains constant and equals 2.1(1) $\mu_B$/atom) between 1.5 and 150 K. Meanwhile, the φ angle slightly increases from 85 to 95 °, rotating toward from **c** to ***a*** axis. A sharp reduction of both Tb and Fe moments is observed around 160 K which corresponds to the cell parameter contraction discussed earlier. Above this temperature, the Tb and Fe magnetic moments continue to slightly decrease up to 220 K, whereas the φ angle increases more sharply above 170 K. Above 220 K, the NPD patterns are refined in a nuclear cell as expected for a paramagnetic state.

The large drop of magnetic moments around 150-160 K in Fig. 6 is in excellent agreement with a, b, c and volume drop occurring at the same temperature range. It is noteworthy that the lattice parameters only exhibit usual thermal lattice expansion above about 200 -220K precisely where the $M$(T) curve presents a bump. This clearly indicates the existence of remaining magnetic contribution above 160 K. This must be connected to the fact that Fig. 1 shows a decrease of the spontaneous magnetization contribution. However, a careful look at Fig. 1e demonstrates the presence of residual spontaneous magnetization contribution which disappears around 200 K. A result in agreement with the thermal evolution of the neutron diffraction intensities as shown in Fig. 5.

Altogether these facts bear witness to the presence of two different temperature domains in the magnetic phase diagram of $TbFe_2D_{4.2}$ in the 2 to 150 K and the second in the range from 150-160 to about 200 K. According to previous results on $YFe_2(H,D)_{4.2}$ the temperature around 200 K corresponds to the ordering driven mainly by the Fe sublattice. We conclude that the transition around 150 K corresponds to the evolution of the Tb sublattice magnetic moment whereas the Fe sublattice keeps essentially the same magnetic moment magnitude from 2 up to 150K. Above this temperature the Phi angle of the moment relative to the c axis evolved away from 90°. This occurs together with a decrease of both Fe and Tb magnetic moments. The observed change of magnetic behavior is most probably associated to the large anisotropic magnetovolume change occurring around 160-170K as shown in Figure 4. The precise knowledge of the magnetic behavior of this compound may deserve further study, for instance on single crystal or using polarized neutron.

It is also noted that the magnetic ordering temperature of $TbFe_2D_{4.2}$ is significantly lower than that of $TbFe_2$ ($T_C$= 695 K) [52]. This decrease may be linked to the weakening of Tb-Fe interactions, as the insertion of deuterium leads to an increase in the interatomic distances between the Fe and Tb atoms.

The previous studies on $YFe_2(H,D)_{4.2}$ compounds have shown that the sharp cell volume contraction was due to a transition from a ferromagnetic to an antiferromagnetic state of the Fe sublattice [39, 40, 47]. Similar observations were reported for $Y_{1-x}Er_xFe_2D_{4.2}$ ($x$ = 0.3 and 0.5) and $Y_{0.9}Tb_{0.1}Fe_2D_{4.2}$ compounds [41, 42, 44]. In all these substituted compounds the magnetic ordering temperature of Er or Tb moments was below the FM-AFM transition, and only the Fe sublattice was ordered at $T_{FM-AFM}$. The AFM magnetic structure displays a $k$ = (0,1/2,0) propagation vector and corresponds to an inversion of the Fe moments direction from one half magnetic cell to the other. The moments are parallel to the ($a$,$c$) plane. The AFM-FM transition was explained by the itinerant electron metamagnetic (IEM) behavior of the Fe sublattice with one Fe site over 8 losing it ordered moment at the transition. Because this Fe site is located at the intersection of two tetrahedra, it induces an antiferromagnetic state more stable than the ferromagnetic one [47]. In the case of $TbFe_2D_{4.2}$, the Tb moment at 150 K is around 2.5 $\mu_B$, large enough to influence the behavior of the Fe sublattice. The exchange coupling between the Fe and Tb magnetic sublattices above 160 K, prevents the ordering of the compound in the AFM structure occurring for Fe sublattice alone as observed for $YFe_2D_{4.2}$.

Therefore, the structure remains ferrimagnetic and only a sharp reduction of the Fe and Tb moments is observed at 160 K. The absence of metamagnetic behavior of the $M(\mu_0 H)$ curves above 160 K -which was observed above $T_{FM-AFM}$ transition temperature for all the other $Y_{1-x}R_xFe_2D_{4.2}$ deuterides and explained by a field induced AFM-FM transition - confirms that a different type of magnetic transition occurs at 160 K. However, the sharp decrease in cell parameters and magnetic moments at 160 K indicates that the Fe sublattice exhibits an IEM behavior at this temperature.

It is also interesting to compare the variation of the transition temperature related to the IEM behavior of the Fe sublattice of $TbFe_2D_{4.2}$ with those of $YFe_2D_{4.2}$ ($T$ = 84 K) and $Y_{0.9}Tb_{0.1}Fe_2D_{4.2}$ ($T$ = 91 K) [44]. Assuming that the difference of transition temperatures between $YFe_2D_{4.2}$ and $YFe_2H_{4.2}$ ($\Delta T$ = 47 K) is purely related to a cell volume expansion ($\Delta V$ = 2.0 Å$^3$) [40], the temperature increase is 22.5 K/Å$^3$ (in the $C2/m$ space group description). The difference of cell volume is 1.595 Å$^3$ and yields a calculated temperature increase $\Delta T$ = 37.5 K between $TbFe_2D_{4.2}$ and $YFe_2D_{4.2}$. The IEM transition temperature for $TbFe_2D_{4.2}$ should be therefore around 121.5 K, for a simple magnetovolume effect. Given that the 38.5 K difference between the experimental (160 K) and calculated (121.5 K) values is due to the influence of the Tb moments affecting this transition temperature, an equivalent applied field of 5.4 T can be estimated using the slope $\Delta\mu_0 H_{Trans}/\Delta T$ = 0.14 T/K measured for the $YFe_2(H,D)_{4.2}$ compounds [40].

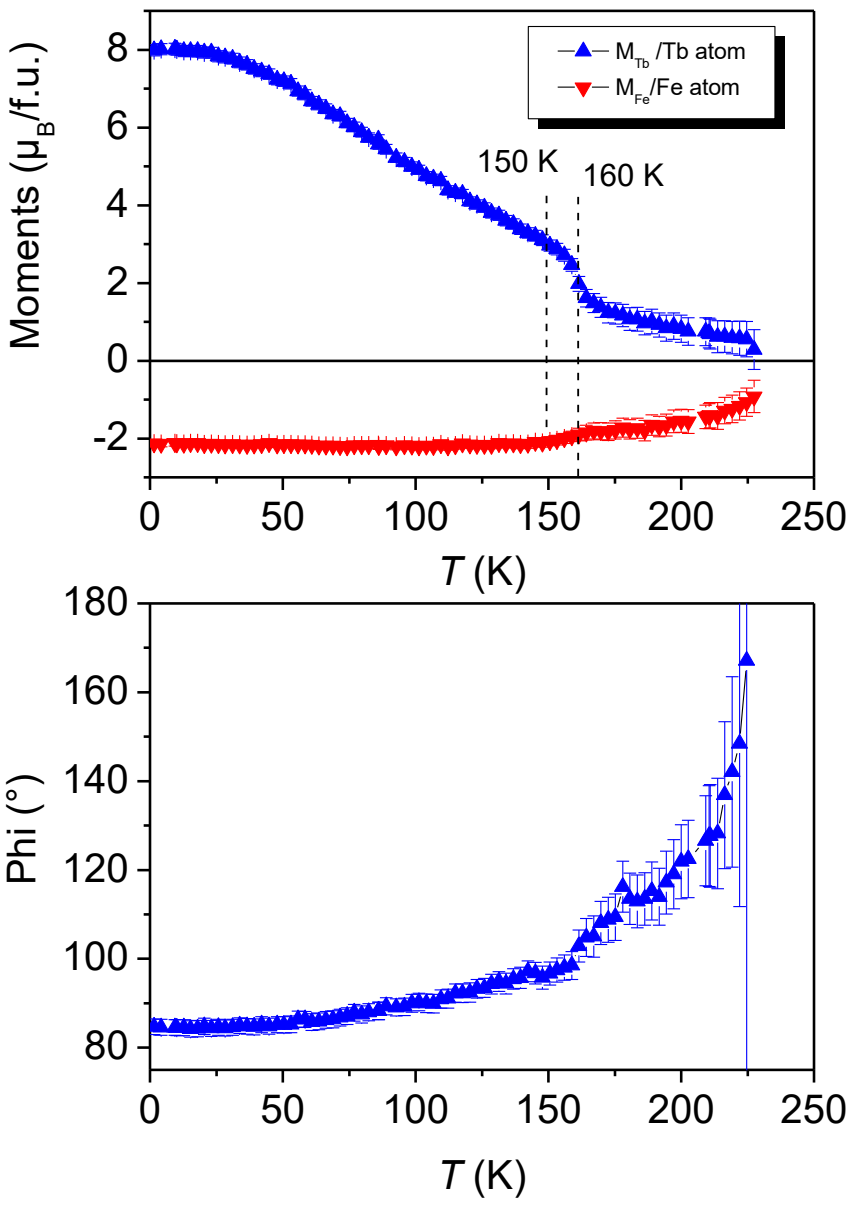

**Figure 6**: Evolution of the Tb and Fe moments as a function of temperature (top) and of the φ angle (bottom) of the moment relative to the ***c*** axis for the deuteride $TbFe_2D_{4.2}$ obtained by the Rietveld refinement of the NPD patterns from D1B.

In contrast, the 7±1 K difference in IEM transition temperature at zero field between $Y_{0.9}Tb_{0.1}Fe_2D_{4.2}$ and $YFe_2D_{4.2}$ is well explained by a simple magnetovolume effect as the 0.34 $Å^3$difference in cell volume yields a temperature difference of 8 K. This is consistent with the lower ordering temperature of Tb (around 40 K) observed in $Y_{0.9}Tb_{0.1}Fe_2D_{4.2}$, indicating that Tb moment has negligible influence at 91 K.

*c) Neutron diffraction measurements versus applied magnetic field*

The NPD patterns, which were measured on E6 diffractometer at HZB, were recorded at 2 K versus an applied field ranging from 0 to 12 T. The intensity of the magnetic peak increases with the applied field (Fig. 7a). Figure 7b shows the patterns at 0 and 12 T and their difference, with indexation of Bragg peaks in *Pc* space group. The variation of the surface of the peak (Fig. 7c) and of the corresponding normalized intensity ($I/I_0$) (Fig. 7d) of the four main magnetic peaks studied before increases versus temperature. The larger relative change in intensity is for the B3 Bragg peak, which is related to only Tb magnetic moment. The second largest change is for the B1 Bragg peak, which contains contributions from both the Tb and Fe magnetic moments, as well as nuclear information. The peak positions are not changing with magnetic field, indicating that the cell parameters remain constant upon applied fields. The refinement of the NPD patterns reveals that both Tb and Fe moments increase versus applied field, with a discontinuity around 3 T in agreement with the magnetization curve at 4.2 K. The angle between the moment and the ***c*** axis increases progressively, indicating a rotation of the moments from ***c*-**axis towards the ***a*-**axis. In another way $M_x$ increases at the expenses of $M_z$ for both Tb and Fe. This can explain the absence of saturation in the $M(\mu_0H)$ curves even at 4.2 K and 50 T, as the structural anisotropy of the monoclinic structure is added to the one related to Tb. Aligning the Fe and Tb moment parallel to the ***a*** axis will require a much larger external field.

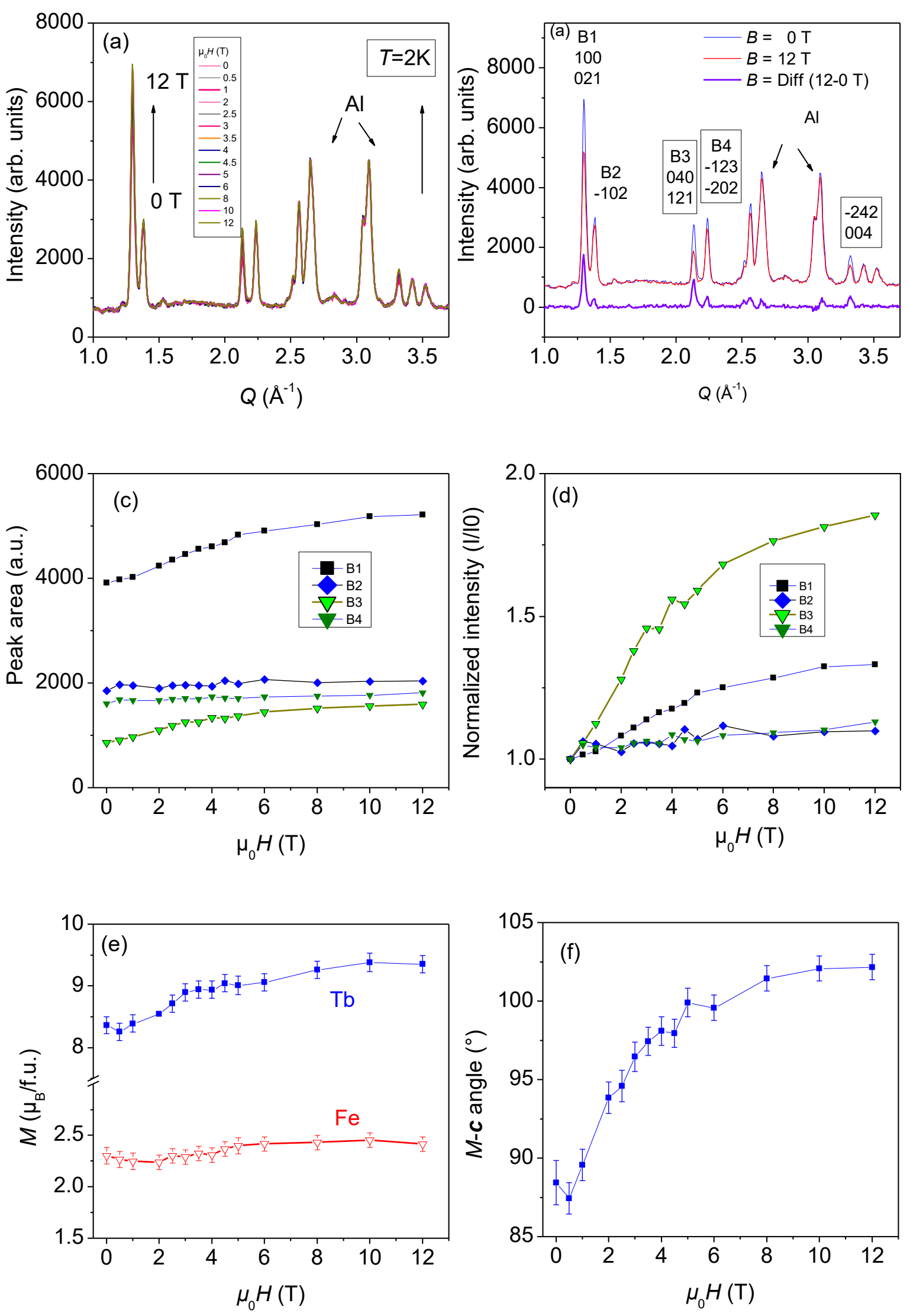

**Figure 7** : Evolution of a) the NPD patterns of $TbFe_2D_{4.2}$ measured on E6 (HZB) at different applied fields at 2 K, b) difference between the 2 K patterns at 0 and 12 T and peak indexation, c) maximum peak intensity versus temperature of the four magnetic peaks at low angle d) Normalized peak intensity $I/I_0$ with $I_0$ the intensity at zero field, e) refined Tb and Fe moments, f) angle between the moment and the ***c*** axis.

The evolution of the NPD patterns measured at 3T versus temperature from 2 K to 160 K in drawn in Figure 8. As for the NPD measurements performed at zero field on D1B, the intensity of the magnetic peak progressively decreases versus temperature. The Tb and Fe moment as well as the angle between the direction of the moment and the ***c*** axis are plotted in Figures 8c and 8d respectively. $M_{Tb}$ progressively decreases from 8.9 to 3.3 $\mu_B$/f.u. and $M_{Fe}$ from 2.3 to 1.8 $\mu_B$/f.u.. This indicates, as expected for a heavy rare-earth that the reduction of the Tb is more sensitive to the temperature changes. An increase of the angle between the magnetic moments and the ***c*** axis increases, i.e. a rotation toward the *a* axis as observed at zero field (Figure 6). Unfortunately, the NPD measurements under 3T were limited to a maximum temperature of 160 K, which do not allow to see a shift of the IEM transition temperature versus applied field.

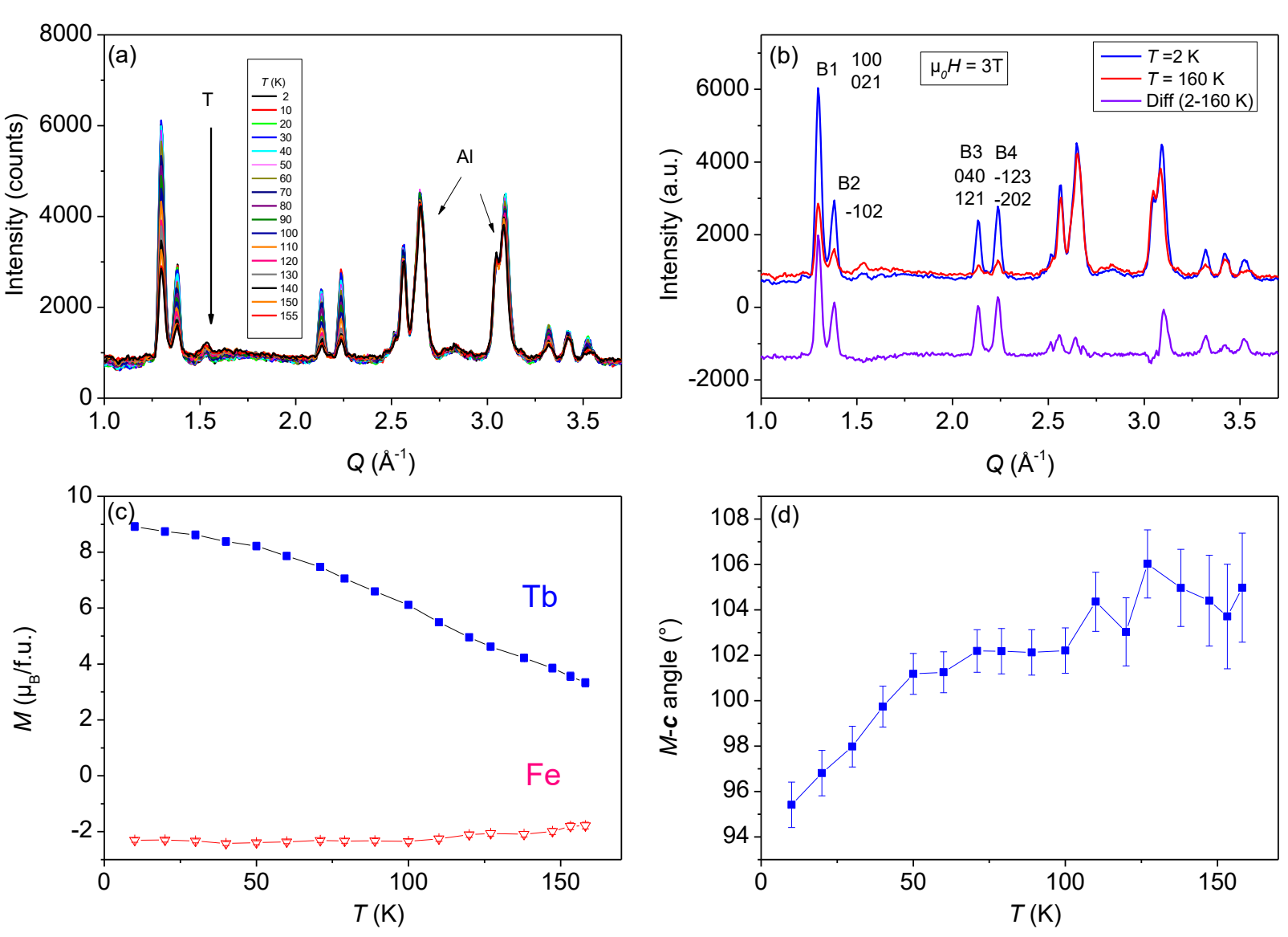

**Figure 8** : Evolution of a) the NPD patterns measured on E6 (HZB) at different temperatures between 2 and 155 K with a field of 3 T, b) difference between the patterns at 2 K and 160 K, c) refined Tb and Fe moments d) angle between the moment and the ***c***-axis.

## Conclusions

In this study, we analyzed the magnetic properties of $TbFe_2D_{4.2}$ using complementary magnetic measurement techniques and neutron diffraction experiments. We observed a ferrimagnetic ground state with $M_{Tb}$= 8.2(2) $\mu_B$ and $M_{Fe}$= 2.1(2) $\mu_B$. The absence of saturation, even at a high magnetic field, reveals the large anisotropy induced by Tb. Upon heating, the Tb moment progressively decreases, while the Fe moment remains constant up to of approximately 150 K150 K. At 160 K, a sharp decrease in both the Tb and Fe moments, as well as in the monoclinic cell parameters, is observed. Magnetization curves indicate a Curie temperature of around 150 K; however, the neutron diffraction pattern shows that Tb and Fe remain ordered at temperatures as high as 220 K. The transition at 160 K shows some similarities with the FM-AFM transition observed in $YFe_2(H,D)_{4.2}$ compounds and in those where Y is partially substituted by Er or Tb. However, the AFM structure could not be observed above 160 K, nor could a field-induced magnetic transition be observed in the $M(\mu_0 H)$ curves. The ferrimagnetic structure is still observed between 160 and 220 K with a decrease in the Fe and Tb moments. This can be attributed to the molecular field induced by the Tb moments above 160 K, which prevents the ordering of the Fe sublattice in the AFM structure. The anisotropic cell parameter variation above 220 K results from the competition between thermal expansion and the reduction of monoclinic distortion.

**Acknowledgements**

We are thankful to T. Leblond and F. Cuevas for their help to measure the NPD patterns at ILL on D1B diffractometer and to K. Provost for participation to NPD measurements on E6 at HZB. We are grateful for the time allocated to perform neutron diffraction experiments. to the Laboratoire Léon Brillouin (3T2 instrument), the CNRS for beam time allocation on D1B instrument at the Institut Laue Langevin (proposal CRG-939) and the Helmholtz Zentrum Berlin (HZB) in Berlin, Germany (E6 instrument, proposal 14100452) for the allocated beam time to perform the neutron experiments. The research project (HZB) has been supported by the European Commission under the 6th Framework Program through the Key Action: Strengthening the European Research Area, Research Infrastructures. Contract n°: RII3-CT-2003-505925 (NMI3). We are grateful to local contacts: Florence Porcher for the measurements on 3T2 diffractometer at LLB, and to Andreas Hoser for the measurements on E6 spectrometer at HZB. We acknowledge the support of the HLD at HZDR, member of the European Magnetic Field Laboratory (EMFL) and the time given to perform the magnetic measurement under pulsed magnetic field (ref. DMA08-217) with the help of Yurii Skourski as local contact for the measurement of $TbFe_2D_{4.2}$ at 4.2 K.

**References**

[1] A.E. Clark, Magnetostrictive rare earth-$Fe_2$ compounds, in: K.H.J. Buschow, E.P. Wohlfarth (Eds.) Handbook of Ferromagnetic Materials, Elsevier, North-Holland, Amsterdam 1980, pp. 531-589.

[2] N.C. Koon, C.M. Williams, B.N. Das, Giant magnetostriction materials, J. Magn. Magn. Mat., 100 (1991) 173-185.https://doi.org/10.1016/0304-8853(91)90819-V

[3] L. Ruiz de Angulo, C.A.F. Manwaring, D.G.R. Jones, J.S. Abell, I.R. Harris, Powder Metallurgical Processing of $Tb_{0.27}Dy_{0.73}Fe_{2-x}$ ($0.5 \geq x \geq 0.1$) from Fine Hydride Powder, Z. Phys. Chemie-Int. J. Res. Phys. Chem. Chem. Phys., 183 (1994) 427-435.https://10.1524/zpch.1994.183.Part_1_2.427

[4] G. Youhui, L.-C. Tai, J.H. Zhu, Magnetic properties and magnetostriction in $Tb_{0.5}Dy_{0.5}(Fe_{0.9}Mn_xAl_{0.1-x})_{1.95}$ compounds, J. Magn. Magn. Mat., 152 (1996) 379-382.https://doi-org.inc.bib.cnrs.fr/10.1016/0304-8853(95)00470-X

[5] Y.G. Shi, S.L. Tang, Y.J. Huang, L.Y. Lv, Y.W. Du, Anisotropy compensation and magnetostriction in $Tb_xNd_{1-x}Fe_{1.9}$ cubic Laves alloys, Appl. Phys. Lett., 90 (2007) 142515.https://doi.org/10.1063/1.2721128

[6] S.N. Jammalamadaka, G. Markandeyulu, K. Balasubramaniam, Magnetostriction and anisotropy compensation in $Tb_xDy_{0.9-x}Nd_{0.1}Fe_{1.93}$ [$0.2<x<0.4$], Appl. Phys. Lett., 97 (2010) 242502.doi:http://dx.doi.org/10.1063/1.3525577

[7] J.J. Liu, R. Wang, H.Y. Ying, X.C. Liu, J. Du, Structural, magnetic and magnetostrictive properties of Co-doped $Tb_{1-x}Ho_xFe_2$ ($0<x<1.0$) alloys, J. Appl. Phys., 110 (2011) 073915.http://dx.doi.org/10.1063/1.3647757

[8] H.Y. Yin, J.J. Liu, Z.B. Pan, X.Y. Liu, X.C. Liu, L.D. Liu, J. Du, P.Z. Si, Magnetostriction of $Tb_xDy_{1-x}Nd_{0.1}(Fe_{0.8}Co_{0.2})_{1.93}$ compounds and their composites ($0.20<x<0.60$), J. Alloys Compds, 582 (2014) 583-587.https://doi-org.inc.bib.cnrs.fr/10.1016/j.jallcom.2013.08.092

[9] A.S. Ilyushin, I.S. Tereshina, N.Y. Pankratov, T.A. Aleroeva, Z.S. Umhaeva, A.Y. Karpenkov, T.Y. Kiseleva, S.A. Granovsky, M. Doerr, H. Drulis, E.A. Tereshina-Chitrova, The phenomenon of magnetic compensation in the multi-component compounds $(Tb,Y,Sm)Fe_2$ and their hydrides, J. Alloys Compds, 847 (2020) 155976.https://doi.org/10.1016/j.jallcom.2020.155976

[10] A.E. Clark, H.T. Savage, Giant magnetically induced changes in elastic-moduli in $Tb_3Dy_7Fe_2$, Ieee Transactions on Sonics and Ultrasonics, SU22 (1975) 50-52.https://doi.org/10.1109/t-su.1975.30775

[11] O.D. McMasters, J.D. Verhoeven, E.D. Gibson, Preparation of Terfenol-D by float zone solidification, J. Magn. Magn. Mat., 54-57 (1986) 849-850.https://doi-org.inc.bib.cnrs.fr/10.1016/0304-8853(86)90281-7

[12] J.B. Restorff, H.T. Savage, A.E. Clark, M. Wun-Folgle, Preisach modeling of hysteresis in Terfenol, J. Appl. Phys., 67 (1990) 5016-5018. https://doi-org.inc.bib.cnrs.fr/10.1063/1.344708

[13] Y. Guo, Y.D. Zhang, H.Y. Su, F.X. Zhu, G. Yi, J.F. Wang, Magnetic-field tuning whispering gallery mode based on hollow microbubble resonator with Terfenol-D-fixed, Applied Optics, 58 (2019) 8889-8893.https://10.1364/ao.58.008889

[14] B.C. Li, T.L. Zhang, C.B. Jiang, J.W. Gu, Low eddy current loss of Terfenol-D/epoxy particulate magnetostrictive composites prepared using the particle phosphatizing treatment method, J. Magn. Magn. Mat., 508 (2020) 166869.https://10.1016/j.jmmm.2020.166869

[15] W.M. Huang, Z.Y. Zhang, P.P. Guo, X.B. Feng, L. Weng, Measurement and calculation for high frequency magnetic losses of Terfenol-D alloy rod under coupled stress and DC bias fields, AIP Adv., 13 (2023) 115013.https://10.1063/5.0175744

[16] K.L. Zhao, P. Pang, J. Luo, B.H. Ma, W. Gao, J.J. Deng, Effects of Annealing on Coercivity and Magnetization of Terfenol-D Thin Films, IEEE Trans. Magn., 59 (2023) 2000104.https://10.1109/tmag.2023.3237937

[17] A.D.M. Charles, A.N. Rider, S.A. Brown, C.H. Wang, The effect of high energy ball milling on Terfenol-D, J. Alloys Compds, 1010 (2025) 178030.https://10.1016/j.jallcom.2024.178030

[18] M.B. Moffett, A.E. Clark, M. Wunfogle, J. Linberg, J.P. Teter, E.A. McLaughlin, Characterization of Terfenol-D for magnetostrictive transducers, J. Acoust. Soc. America, 89 (1991) 1448-1455.https://doi.org/10.1121/1.400678

[19] D. Satpathi, J.A. Moore, M.G. Ennis, Design of a Terfenol-D based fiber-optic current transducer, IEEE Sens. J., 5 (2005) 1057-1065.https://doi.org/10.1109/jsen.2005.850996

[20] M.A. Patil, R. Kadoli, Engineers guide to Terfenol-D actuators: Design, performance, and real-world applications, Sensors and Actuators Reports, 8 (2024) 100236.https://10.1016/j.snr.2024.100236

[21] S. Ranjbar, F. Nazari, R. Hajizadeh, A magnetically switchable demultiplexer via Terfenol-D in phononic crystal, J. Magn. Magn. Mat., 609 (2024) 172484.https://10.1016/j.jmmm.2024.172484

[22] E.M. Omrani, F. Nazari, Magnetically tunable 4 x 2 encoder utilizing Terfenol-D-embedded phononic crystal ring resonators, Journal of Science-Advanced Materials and Devices, 10 (2025) 100861.https://10.1016/j.jsamd.2025.100861

[23] Z. Yang, S.L. Pu, T.F. Xu, W.N. Liu, C.C. Zhang, M. Lahoubi, Fiber-Optic Magnetic Field Sensor Based on Curved Terfenol-D Rod Combined With FBG, IEEE Sens. J., 25 (2025) 6234-6241.https://10.1109/jsen.2025.3525539
[24] K.Q. Ye, Y.L. Xu, J.Q. Zheng, Real-time online resonance frequency tuning of piezoelectric ultrasonic transducer through nesting Terfenol-D into composite horn, Rev. Sci. Instrum., 96 (2025).https://10.1063/5.0235900
[25] L. Ruiz de Angulo, J.S. Abell, I.R. Harris, Influence of hydrogen on the magnetic properties of Terfenol-D, J. Appl. Phys., 76 (1994) 7157-7159.https://10.1063/1.357993
[26] Y. Berthier, T. de Saxce, D. Fruchart, P. Vulliet, Magnetic interactions and structural properties of ternary hydrides $TbFe_2H_x$, Physica B, 130B (1985) 520-523 https://doi.org/10.1016/0378-4363(85)90293-1
[27] D. Fruchart, Y. Berthier, T. Desaxce, P. Vulliet, Structural and magnetic studies of cubic and rhombohedral forms of $ErFe_2H_x$ and $TbFe_2H_x$, J. Solid State Chem., 67 (1987) 197-209.https://doi.org/10.1016/0022-4596(87)90355-0
[28] S.K. Kulshreshtha, O.D. Jayakumar, R. Sasikala, Hydrogen-induced spin reorientation in $TbFe_2H_x$ system, J. Magn. Magn. Mat., 117 (1992) 33-37.https://doi.org/10.1016/0304-8853(92)90288-y
[29] N.V. Mushnikov, N.K. Zajkov, V.S. Gaviko, Inversion of the magnetization near the compensation point in $TbFe_2H_x$ hydrides, J. Alloys Compds, 191 (1993) 63-66.https://doi.org/10.1016/0925-8388(93)90271-n
[30] A.Y. Yermakov, N.V. Mushnikov, N.K. Zajkov, V.S. Gaviko, V.A. Barinov, Magnetic and magnetoelastic properties of amorphous and crystalline $TbFe_2H_x$ hydrides, Philos. Mag. B-Phys. Condens. Matter Stat. Mech. Electron. Opt. Magn. Prop., 68 (1993) 883-890.https://doi.org/10.1080/13642819308217945
[31] N.V. Mushnikov, N.K. Zaikov, V.S. Gaviko, A.V. Korolev, A.E. Ermakov, Hydrogen-induced magnetic anisotropy and magnetostriction in rare-earth intermetallics with $MgCu_2$-type structure, Russian Metallurgy, (1995) 86-91.https://doi.org/10.1134/S0031918X13120077
[32] N.K. Zaykov, N.V. Mushnikov, A.Y. Yermakov, Magnetocrystalline anisotropy and magnetostriction of $TbFe_2H_x$ hydride, Fiz. Metallov Metalloved., 79 (1995) 50-60
[33] N.K. Zaikov, N.V. Mushnikov, V.S. Gaviko, Reversible TbFe2Hx-base permanent magnet, Russian Metallurgy, (1996) 88-91

[34] N.K. Zajkov, N.V. Mushnikov, V.S. Gaviko, A.Y. Yermakov, Effect of high-temperature hydrogen treatment on magnetic properties and structure of $TbFe_2$-based compounds, Int. J. Hydr. Energ., 22 (1997) 249-253.https://doi.org/10.1016/s0360-3199(96)00168-1
[35] V. Paul-Boncour, O. Isnard, Structural transitions related to order-disorder and thermal desorption of D atoms in $TbFe_2D_{4.2}$, J. Alloys Compds, 1058 (2026) 187024.https://doi.org/10.1016/j.jallcom.2026.187024
[36] N.K. Zaikov, N.V. Mushnikov, V.S. Gaviko, A.E. Ermakov, Magnetic properties and structure of hydrogen-amorphized intermetallic compounds $RFe_2H_x$ (R=Y, Gd, Tb, Dy, Ho, Er), Phys. Solid State, 39 (1997) 810-814.https://doi.org/10.1134/1.1129975
[37] T. Leblond, V. Paul-Boncour, M. Guillot, O. Isnard, Metamagnetic transitions in $RFe_2(H,D)_{4.2}$ compounds (R=Y, Tb), J. Appl. Phys., 101 (2007) 09G514.https://doi.org/10.1063/1.2710456
[38] V. Paul-Boncour, T. Mazet, Investigation of compounds for magnetocaloric applications: $YFe_2H_{4.2}$, $YFe_2D_{4.2}$ and $Y_{0.5}Tb_{0.5}Fe_2D_{4.2}$, J. Appl. Phys., 105 (2009) 013914 https://doi.org/10.1063/1.3055348
[39] O. Isnard, V. Paul-Boncour, Z. Arnold, On the origin of the giant isotopic effect of hydrogen on the magnetic properties of $YFe_2A_{4.2}$ (A = H, D): A high pressure study, Appl. Phys. Lett., 102 (2013) 122408.

https://doi.org/10.1063/1.4798260
[40] V. Paul-Boncour, M. Guillot, G. Wiesinger, G. André, Giant isotopic effect on the itinerant-electron metamagnetism in $YFe_2(H_yD_{1-y})_{4.2}$, Phys. Rev. B, 72 (2005) 174430.https://10.1103/PhysRevB.72.174430
[41] V. Paul-Boncour, M. Guillot, O. Isnard, A. Hoser, High field induced magnetic transitions in the $Y_{0.7}Er_{0.3}Fe_2D_{4.2}$ deuteride, Phys. Rev. B, 96 (2017) 104440.https://doi.org/10.1103/PhysRevB.96.104440
[42] V. Paul-Boncour, O. Isnard, M. Guillot, A. Hoser, Metamagnetic transitions in $Y_{0.5}Er_{0.5}Fe_2D_{4.2}$ deuteride studied by high magnetic field and neutron diffraction experiments, J. Magn. Magn. Mat., 477 (2019) 356-365.https://doi.org/10.1016/j.jmmm.2019.01.056
[43] V. Paul-Boncour, O. Isnard, V. Shtender, Y. Skourski, M. Guillot, Origin of the metamagnetic transitions in $Y_{1-x}Er_xFe_2(H,D)_{4.2}$ compounds, J. Magn. Magn. Mat., 512 (2020) 167018.https://doi.org/10.1016/j.jmmm.2020.167018

[44] V. Paul-Boncour, V. Shtender, K. Provost, M. Phejar, F. Cuevas, Y. Skourski, O. Isnard, Origin of the metamagnetic transitions in $Y_{0.9}Tb_{0.1}Fe_2D_{4.3}$, J. Solid State Chem., 338 (2024) 9.https://doi.org/10.1016/j.jssc.2024.124898
[45] O. Isnard, V. Paul-Boncour, Z. Arnold, C.V. Colin, T. Leblond, J. Kamarad, H. Sugiura, Pressure-induced changes in the structural and magnetic properties of $YFe_2D_{4.2}$, Phys. Rev. B, 84 (2011) 094429.https://doi.org/10.1103/PhysRevB.84.094429
[46] Z. Arnold, O. Isnard, V. Paul-Boncour, Influence of high pressure on the remarkable itinerant electron behavior in $Y_{0.7}Er_{0.3}Fe_2D_{4.2}$ compound, J. Appl. Phys., 133 (2023) 173901.https://doi.org/10.1063/5.0141855
[47] V. Paul-Boncour, M. Guillot, O. Isnard, B. Ouladdiaf, A. Hoser, T. Hansen, N. Stuesser, Interplay between crystal and magnetic structures in $YFe_2(H_\alpha D_{1-\alpha})_{4.2}$ compounds studied by neutron diffraction, J. Solid State Chem., 245 (2017) 98-109.http://dx.doi.org/10.1016/j.jssc.2016.09.002
[48] J. Rodriguez-Carvajal, Recent Advances in Magnetic Structure Determination by Neutron Powder Diffraction, Physica B, 192 (1993) 55-69.https://doi-org.inc.bib.cnrs.fr/10.1016/0921-4526(93)90108-I
[49] D.G. Westlake, Site occupancies and stoichiometries in hydrides of intermetallic compounds : geometric considerations, J. Less-Common Met., 90 (1983) 251-273
[50] A.C. Switendick, Band structure calculations for metal hydrogen systems, Z. Phys. Chem. Neue Fol., 117 (1979) 89-112.https://doi.org/10.1524/zpch.1979.117.117.089
[51] J. Ropka, R. Cerný, V. Paul-Boncour, Local deuterium order in apparently disordered Laves phase deuteride $YFe_2D_{4.2}$, J. Solid State Chem., 184 (2011) 2516-2524.https://doi.org/10.1016/j.jssc.2011.07.028
[52] Y.J. Tang, Transition-metal substitution effect on magnetic and magnetostrictive properties of $TbFe_2$ compounds, J. Magn. Magn. Mat., 167 (1997) 245-248.https://doi.org/10.1016/S0304-8853(96)00331-9